\documentclass[aps,prl,twocolumn,showpacs,superscriptaddress,groupedaddress]{revtex4}  
\usepackage{graphicx}  
\usepackage{dcolumn}   
\usepackage{bm}        
\usepackage{amssymb}   

\usepackage{graphicx}
\usepackage{dcolumn}
\usepackage{bm}
\usepackage[utf8]{inputenc}
\usepackage{amsmath}
\usepackage{amsthm}
\usepackage{amsmath}
\usepackage{amsfonts}
\usepackage{enumitem}
\usepackage{bibunits}
\usepackage[multiple]{footmisc}
\usepackage{graphicx}
\usepackage{framed} 
\usepackage{epstopdf}
\usepackage{setspace}
\usepackage{amssymb}

\newtheorem{lem}{Lemma}

\begin{document}

\widetext
%

\leftline{Comment to {\tt amirsagiv@mail.tau.ac.il}}


\title{Loss of phase and universality of stochastic interactions between laser beams}

\author{Amir Sagiv, Adi Ditkowski, Gadi Fibich \\ %
Department of Applied Mathematics, Tel Aviv University, Tel Aviv 6997801, Israel}

\begin{abstract}
We show that all laser beams gradually lose their initial phase information in nonlinear propagation. Therefore, if two beams travel a sufficiently long distance before interacting, it is not possible to predict whether they would intersect in- or out-of-phase. Hence, if the underlying  propagation model is non-integrable, deterministic predictions and control of the interaction outcome become impossible. Because the relative phase between the two beams becomes uniformly distributed in $[0,2\pi]$, however, the statistics of the interaction outcome are universal, and can be efficiently computed using a polynomial-chaos approach, even when the distributions of the noise sources are unknown.
\end{abstract}

\pacs{42.65.Tg, 42.65.Sf,42.65.Jx}
\maketitle

Nonlinear interactions between two or more laser beams, pulses, and filaments \cite{segev1999optical} are related to applications ranging from modulation methods in optical communication \cite{skidin2016mitigation}, to coherent combination of beams~\cite{fan2005laser, daniault2015xcan, mourou2014fiber, mourou2013future, bellanger2010collective}, interactions between filaments in atmospheric propagation \cite{couairon2007femtosecond} and ignition of nuclear fusion using up to 192 beams \cite{kruer1996energy}. In the integrable one-dimensional cubic case, such interactions can only lead to phase and lateral shifts, which can be computed analytically using the Inverse Scattering Transform~\cite{zakharov1973interaction, ablowitz2004NLS, mitschke1987experimental}. In the non-integrable case, however, richer dynamics are possible, including beam repulsion, breakup, fusion and spiraling~\cite{segev1999optical, snyder1993collisions, Break2001Fusion, tikhonenko1996three}. Since the outcome of the interaction strongly depends on the relative phases of the beams \cite{garcia1997observation}, one can use the initial phase to control the interaction dynamics~\cite{ishaaya2007self}. 
Nonlinear interactions between solitary waves were also studied in other physical systems \cite{scott1973soliton, kivshar1989dynamics}, such as fiber optics \cite{agrawal2007nonlinear, gordon1983interaction}, waveguide arrays \cite{meier2004discrete}, water waves \cite{craik1988wave, su1980head}, plasma waves~\cite{zabusky1965interaction} and Bose-Einstein condensates~\cite{nguyen2014collisions}.
 
In previous studies it was shown, both theoretically and experimentally, that when a laser beam undergoes an optical collapse, its initial phase is "lost", in the sense that the small shot-to-shot variations in the input beam lead to large changes in the nonlinear phase shift of the collapsing beam~\cite{fibich2011continuations,Merle-92}. Therefore, if two such beams intersect after they collapsed, one cannot predict whether they will intersect in- or out-of-phase, and so post-collapse interactions between beams become "chaotic" and cannot be controlled~\cite{gaeta2012loss}. Loss of phase can also interfere with imaging in nonlinear medium~\cite{goy2013imaging, barsi2009imaging}. Note that loss of phase does not imply a loss of coherence, but rather that at any given propagation distance, the coherent beam can only be determined up to an unknown constant phase. 

In this study we show that {\em loss of phase is ubiquitous in nonlinear optics}. Thus, while collapse accelerates the loss of phase process, non-collapsing or mildly-collapsing beams can also undergo a loss of phase. The loss of phase builds up gradually with the propagation distance, i.e., the shot-to-shot variations of the beam's nonlinear phase shift increase with the propagation distance, and approach a uniform distribution in $[0,2\pi]$ at sufficiently large distances. As noted, because of the loss of phase, deterministic predictions and control of interactions between laser beams become impossible. We show, however, that loss of phase allows for accurate predictions of the statistical properties of these stochastic interactions, even without any knowledge of the noise source and characteristics. Indeed, because the relative phase between the beams becomes uniformly distributed in~$[0,2\pi]$, the statistics of the interaction are universal, and can be computed using a {\em "universal model"} in which the {\em only} noise source is a uniformly distributed phase difference between the input beams. These computations can be efficiently performed using a polynomial-chaos based approach.

The propagation of laser beams in a homogeneous medium is governed by the dimensionless nonlinear Schr{\"o}dinger equation (NLS) in $d+1$ dimensions
\begin{equation}\label{eq:nls_gen}
i\frac{\partial}{\partial z} \psi (z,{\bf x})  + \nabla ^2 \psi  + N(|\psi|)\psi =0  \, ,
\end{equation}
where $z$ is the propagation distance, ${\bf x} = (x_1,\ldots ,x_d)$ are the transverse coordinates (and/or time in the anomalous regime), and $\nabla ^2 = \partial ^2 _{x_1} + \cdots + \partial ^2 _{x_d}$ \cite{fibich2015nonlinear}. Here we consider nonlinearities that support stable solitary waves $\psi = e^{i\kappa z}R_{\kappa }({\bf x})$, such as the cubic-quintic NLS
\begin{equation}\label{eq:qc_nls}
i\frac{\partial}{\partial z}\psi (z,{\bf x} ) +\nabla ^2 \psi + |\psi |^{2}\psi  - \epsilon |\psi | ^{4}  \psi	 = 0  \,  , 
\end{equation} 
or the saturated NLS
\begin{equation}\label{eq:sat_nls}
i\frac{\partial}{\partial z}\psi (z,{\bf x}) +\nabla ^2 \psi +  \frac{ |\psi|^2}{1 + \epsilon |\psi|^2} \psi = 0 \,  .
\end{equation}

In a physical system the input beam changes from shot to shot. To model this, we write
\begin{equation}
\psi(z=0, {\bf x};\alpha ) = \psi _0 ({\bf x};\alpha) \, ,
\end{equation}
where $\alpha$ is the noise realization. We denote by $\varphi (z;\alpha) :={\rm arg} \, \psi(z ,x=~0;\alpha) $ the cumulative on-axis phase at $z$, and study the evolution (in~$z$) of the probability distribution function (PDF) of the non-cumulative on-axis phase 
\begin{equation}
\tilde{\varphi}(z;\alpha) :\,= \varphi(z;\alpha) \, {\rm mod} (2\pi) \, .
\end{equation}

\begin{figure}[h!]
\centering
{\includegraphics[scale=0.5]{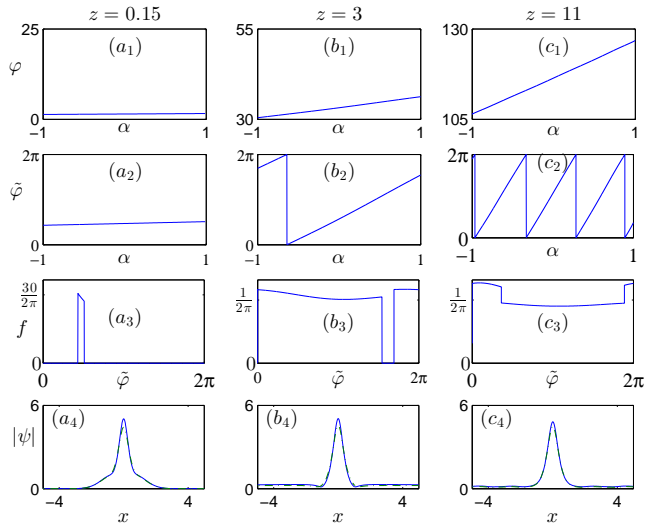}}
\caption{The cubic-quintic NLS \eqref{eq:qc_nls} with $d=1$, $\epsilon =~10^{-3}$, and the initial condition~\eqref{eq:gaus_rnd_ic} at ($a_1$)--($a_4$) $z=~0.15$, ($b_1$)--($b_4$) $z=3$, and ($c_1$--$c_4$) $z=11$. ($a_1$)--($c_1$)~Cumulative on-axis phase as a function of $\alpha$. ($a_2$)--($c_2$) Non-cumulative on-axis phase. ($a_3$)--($c_3$) The PDF of $\tilde{\varphi}$. ($a_4$)--($c_4$) Transverse profile for $\alpha=~1$~(solid) and $\alpha =-1$ (dot-dash).}
\label{fig:lop_gpc}
\end{figure}

For example, consider the one-dimensional cubic-quintic NLS \eqref{eq:qc_nls} with the Gaussian initial condition with a random power
\begin{equation}\label{eq:gaus_rnd_ic}
\psi _0(x;\alpha ) = 3.4(1+0.1\alpha)e^{-x^2}\,   , \quad  \alpha \sim U(-1,1) \, ,
\end{equation}
where $U(-1,1)$ is the uniform distribution in $(-1,1)$. {\em Here we consider the one-dimensional case to emphasize that loss of phase and "chaotic" interactions are not limited to collapsing beams}, as was implied by earlier studies \cite{fibich2011continuations}. At $z=0.15$, the maximal variation of the cumulative phase $\Delta \varphi  := \varphi(\alpha =1 )-\varphi(\alpha  = -1)$ is fairly small~($\approx~ 0.08\pi $), see Fig. \ref{fig:lop_gpc}($a_1$). The corresponding non-cumulative on-axis phase $\tilde{\varphi}$ is identical (Fig. \ref{fig:lop_gpc}($a_2$)), and so the probability distribution function (PDF) of $\tilde{\varphi}$, denoted by $f(\tilde{\varphi})$, is fairly localized (Fig. \ref{fig:lop_gpc}($a_3$)). As the beam continues to propagate ($z=3$), the maximal variation of the cumulative phase increases to $\Delta \varphi \approx 1.8 \pi$, and so $\tilde{\varphi}$ attains most values in $\left[0,2\pi\right]$, though not with the same probability (Fig.~\ref{fig:lop_gpc}($b_1$)--($b_3$)). At an even larger propagation distance ($z=11$), the maximal phase variation  is $\Delta \varphi \approx 6.5\pi$, i.e., slightly over three cycles of $\tilde{\varphi}$, see Fig. \ref{fig:lop_gpc}($c_1$)--($c_2$). At this stage $\tilde{\varphi}$ is nearly uniformly distributed in~$[0,2\pi]$, see Fig. \ref{fig:lop_gpc}($c_3$), which implies that the beam "lost" its initial phase $\varphi (z=0)$. By "loss of phase" we mean that one cannot infer from $f$, the PDF of $\tilde{\varphi}(z;\alpha)$, or from several realizations $\left\{ \varphi (z;\alpha _j \right\}_{j=1}^J$, whether the initial condition was $\psi _0 (x)=c(\alpha)e^{-x^2}$ or $\psi _0 (x)= e^{i\theta }c(\alpha)e^{-x^2}$ for some $0< \theta < 2\pi$. Loss of phase is not accompanied by a "loss of amplitude". Indeed, the differences between the profiles for $\alpha = \pm 1$ remain small throughout the propagation (Fig.~\ref{fig:lop_gpc}($a_4$)--($c_4$)).

%
%
\begin{figure}[h!]
\centering
{\includegraphics[scale=0.5]{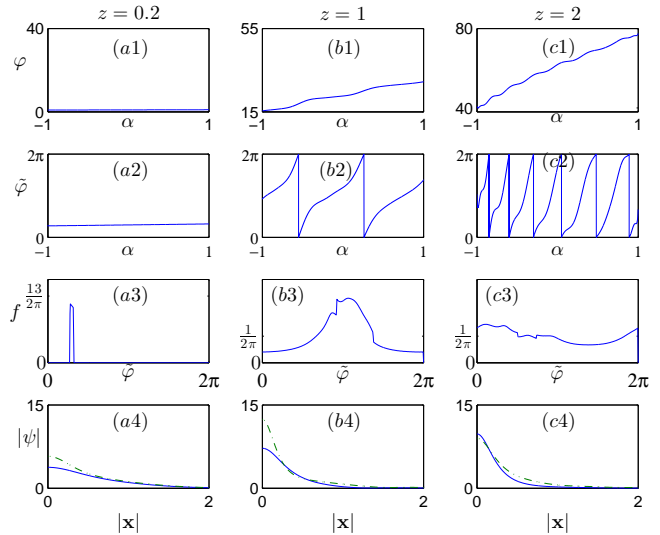}}
\caption{Same as figure \ref{fig:lop_gpc} for $d=2$.}
\label{fig:lop2d}
\end{figure}

\begin{figure}[h!]
\centering
{\includegraphics[scale=0.4]{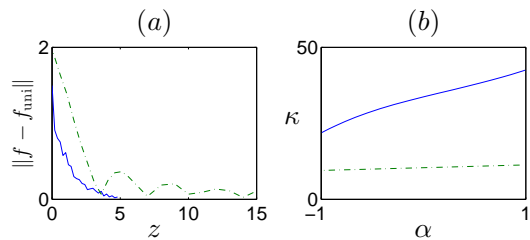}}
\caption{The cubic-quintic NLS~\eqref{eq:qc_nls} with  $\epsilon = 10^{-3}$, and the initial condition~\eqref{eq:gaus_rnd_ic} in one~(dot-dash) and two (solid) dimensions. (a) Distance between the PDF of $\tilde{\varphi}$ and the uniform distribution on $[0,2\pi]$ . (b) The propagation constant of the beam core, see \eqref{eq:steady_rad}, as a function of $\alpha$.}
\label{fig:1d2d}
\end{figure}

We obtained similar results for the same equation and initial condition in two dimensions, see Fig. \ref{fig:lop2d}. To compare the rates at which the PDFs of $\tilde{\varphi}$ converge to the uniform distribution $f _{\rm uni} (y) :\, \equiv \frac{1}{2\pi}$ on $[0,2\pi]$, we plot in Fig. \ref{fig:1d2d}(a) the distance $\| f - f _{\rm uni}  \| :\,=  \int\limits_{0}^{2\pi} \left| f (y) -\frac{1}{2\pi} \right| \, dy$.
The convergence is much faster for $d=2$ than for $d=1$, for reasons that will be clarified later~\footnote{The convergence is not monotone, because the distance has a local minimum whenever $\Delta \varphi = 2\pi k$ for an integer~$k$.}.



\begin{figure}[h!]
\centering
{\includegraphics[scale=0.4]{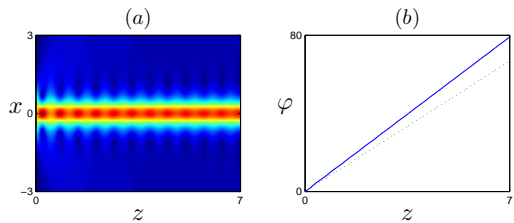}}
\caption{Same as Fig. \ref{fig:lop_gpc}. (a)~The intensity $|\psi(z,x) |^2$ for $\alpha =0$. (b) The on-axis phase for $\alpha  =1$ (solid) and $\alpha =-1$~(dots).}
\label{fig:singleBeamExamples}
\end{figure}

To understand the emergence of loss of phase in Fig.~\ref{fig:lop_gpc}--\ref{fig:lop2d}, we note that after an initial transient, the beam core evolves into a stable solitary wave, see e.g., Fig.~\ref{fig:singleBeamExamples}(a), and so
\begin{equation}\label{eq:steady_rad}
\psi(z,{\bf x};\alpha ) \approx e^{i\eta _0 (\alpha)} e^{i\kappa z} R_{\kappa} ({\bf x}) + {\rm radiation} \, ,
\end{equation}
where $\eta _0 (\alpha)$ is the on-axis phase which is accumulated during the initial transient, $\kappa$ is the propagation constant, and $R_{\kappa}$ is the positive solution of $\nabla ^2 R_{\kappa} -\kappa R_{\kappa}({\bf x})  +N(R_{\kappa}) R_{\kappa} = 0$. By \eqref{eq:steady_rad},
\begin{equation}\label{eq:phase_lin}
\varphi (z;\alpha) \approx  \eta _0 (\alpha) + z\kappa (\alpha )  \, .
\end{equation}  
Thus, the nonlinear phase shift grows linearly with~$z$ at the rate of $\kappa = \kappa (\alpha)$, see figure~\ref{fig:singleBeamExamples}(b).


Since $\alpha$ is randomly distributed, then so is $\kappa (\alpha)$. More generally, \textit{for any initial noise, the beam core evolves into a solitary wave with a random propagation constant $ \kappa (\alpha)$, and so $\varphi$ is given by \eqref{eq:phase_lin}} \footnote{This is also the case with multi-parameter noises, e.g., when the amplitude, the phase and the tilt angle are all random.}. Consequently, the initial on-axis phase is completely lost as $z\to \infty$:

\begin{lem}\label{lem:phase_uni}
Let $\alpha$ be a random variable which is distributed in $[\alpha _{\min},\alpha _{\max}]$ with an absolutely-continuous measure $d\mu $, let $\kappa (\alpha)$ be a continuously differentiable, piece-wise monotone function on $[\alpha _{\min},\alpha _{\max}]$, let $\eta_0(\alpha)$ be continuously differentiable on $[\alpha _{\min},\alpha _{\max}]$, and let~$\varphi$ be given by \eqref{eq:phase_lin}. Then~$\lim\limits_{z\to \infty}  \varphi (z;\alpha)  \, {\rm mod} \, (2\pi) \sim~U([0,2\pi])\, .$

\end{lem}
\textit{Proof:} see SM.

Lemma \ref{lem:phase_uni} provides a new road to the emergence of loss of phase. Indeed, in previous studies \cite{fibich2011continuations, gaeta2012loss,Merle-92}, the loss of phase was caused by the large self-phase modulations (SPM) that accumulate during the initial beam collapse (i.e., by the variation of $\eta _0$ in $\alpha$). Briefly, when a beam undergoes collapse, then in the absence of a collapse-arresting mechanism, $\varphi _0 (\alpha) \to \infty$ as $z\to Z_c (\alpha)$, where $Z_c$ is the collapse point \cite{fibich2011continuations}. 
Therefore, if a beam undergoes a considerable self-focusing before its collapse is arrested, then it accumulates significant SPM, i.e., $\eta _0 (\alpha) \gg 2\pi$. In that case, although small changes in $\alpha$ lead to small relative changes in $\eta _0 (\alpha)$, those are $O(1)$ absolute changes in $\eta _0 (\alpha)$. In this study, however, we consider non-collapsing beams of the 1D NLS, or mildly-collapsing beams of the 2D NLS. Therefore $\Delta \eta _0:=~\eta _0 (\alpha _{\max}) -~\eta _0 (\alpha _{\min} ) \ll~2\pi$. In such cases, the loss of phase builds up gradually with the propagation distance~$z$, and not abruptly during the initial collapse, as in the previous studies.

The loss of phase is a nonlinear phenomenon. Indeed, in the linear propagation regime, $\psi(z,{\bf x}) =(2iz)^{-\frac{1}{2}}e^{u\frac{|{\bf x}|^2}{4z}} \ast \psi _0({\bf x})$. Therefore if $\psi _0 (\alpha _1) - \psi _0 (\alpha_2) \ll 1$ then $\psi (\alpha _1 ) -\psi (\alpha _2) \ll 1$ as well. 

Lemma \ref{lem:phase_uni} is reminiscent of classical results in ergodic theory of irrational rotations of the circle \cite{walters2000ergodic}. Unlike these results, however, Lemma \ref{lem:phase_uni} does not describe the trajectory of a single point on the circle under consecutive discrete phase additions, but rather the convergence of a continuum of points under with continuous linear change with a varying rate $\kappa$.

By \eqref{eq:phase_lin}, the maximal difference in the cumulative phase between solutions grows linearly in~$z$, i.e.,  $$ \Delta \varphi (z) :\,= \varphi (z;\alpha _{\max}) - \varphi (z;\alpha _{\min})\approx \Delta \eta _0  +z \cdot \Delta \kappa   \, , $$
where $\Delta \kappa :\,= \kappa (\alpha _{\max} ) - \kappa (\alpha _{\min} ) $ is the maximal variation in the propagation constant, induced by the noise. As suggested by the proof of Lemma \ref{lem:phase_uni} and by Figs. \ref{fig:lop_gpc} and \ref{fig:lop2d}, $\tilde{\varphi}$~is close to be uniformly distributed once $\Delta \varphi (z)  \gg~2\pi $. Therefore, the characteristic distance for loss of phase is
\begin{equation}\label{eq:nondim_zlop}
Z _{\rm lop} := \frac{2\pi}{\Delta \kappa} \,  .
\end{equation}

Typically, $\Delta \kappa$ is much larger in 2d than in 1d. For example, in Fig. \ref{fig:1d2d}(b) $\Delta \kappa \approx 25$ in 2d, and $\Delta \kappa \approx 1.8$ in  1d. Intuitively, this is because the input beam evolves into a solitary wave, and over a given power range, the propagation constant of the solitary wave changes considerably less in 1d than in 2d~\footnote{Denote the solitary-wave power by $P(\kappa):=~\int|R_{\kappa} | ^2 \, d{\bf x}$. When $\epsilon =0$ in \eqref{eq:qc_nls} or \eqref{eq:sat_nls}, then $\frac{dP}{d\kappa} =0$ for $d=2$ but $\frac{dP}{d\kappa} >0$ for $d=1$\cite{fibich2015nonlinear}. Hence, if $0<\epsilon \ll 1$, then $\frac{dP}{d\kappa}=O(\epsilon)$ for $d=2$, but $\frac{dP}{d\kappa}=O(1)$ for $d=1$. Therefore $\frac{d\kappa}{dP}=O(1)$ for $d=1$ but $\frac{d\kappa}{dP}=O(\frac{1}{\epsilon})$ for $d=2$.}. Hence, by \eqref{eq:nondim_zlop}, the loss of phase occurs much faster in the two-dimensional case than in the one-dimensional case, thus explaining Fig. \ref{fig:1d2d}(a).

\begin{figure}[h!]
\centering
{\includegraphics[scale=0.45]{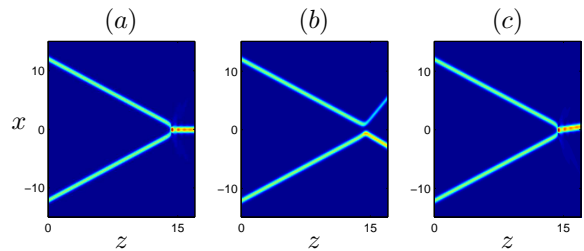}}
\caption{The 1d cubic-quintic NLS~\eqref{eq:qc_nls} with $\epsilon =~2\cdot~10^{-2}$ and the initial condition \eqref{eq:2sol_beams} with $\kappa _0 = 8$. (a) $\kappa _1 = 8$, $\eta_0 = 0$. (b) $\kappa _1 = 8.1$, $\eta_0 = 0$. (c)~$\kappa _1 = 8.1$, $\eta_0 \approx- 0.48\pi$.}
\label{fig:s1q2_interaction}
\end{figure}

A priori, the loss of initial phase has no physical implications, since the NLS~\eqref{eq:nls_gen} is invariant under the transformation $\psi \to e^{i\beta}\psi$. When the NLS \eqref{eq:nls_gen} is non-integrable, however, the \textit{relative phase} between two beams \cite{gaeta2012loss, garcia1997observation, krolikowski1997fusion, tikhonenko1996three} or condensates \cite{nguyen2014collisions} can have a dramatic effect on their interaction, thus making the loss of initial phase physically important. To illustrate that, consider again the cubic-quintic NLS~\eqref{eq:qc_nls}~for $d=1$ with the two crossing beams initial condition
\begin{equation}\label{eq:2sol_beams}
\psi_0 (x) = e^{i\theta x} R_{\kappa _0} (x-a) + e^{i\eta _0}e^{-i\theta x} R_{\kappa_1} (x+a) \, ,
\end{equation}
where $a=12$, $\theta = \frac{\pi}{8}$, $\kappa _0 = 8$, and $R_{\kappa}$
is the solitary wave of  \eqref{eq:qc_nls}.
%
By Galilean invariance, before the beams intersect at $(z_{\rm cross},x_{\rm cross})\approx (14.7,0)$, each beam propagates as a solitary wave, and so
$$ \psi (z,x) \approx e^{i\kappa _0 z}e^{i\theta x-i\theta ^2 z} R_{\kappa _0} (x-a- 2\theta z )  \, $$ $$\qquad\qquad \qquad \quad +e^{i\eta _0} e^{i\kappa _1 z}e^{-i\theta x-i\theta ^2 z} R_{\kappa_1} (x +a+ 2\theta z ) \, .$$
Hence, the difference between the on-axis phases of the two beams at $(z_{\rm cross},x_{\rm cross})$ is
\begin{equation}\label{eq:phase_dif}
\Delta \varphi  \approx (\kappa _1 - \kappa _0)z_{\rm cross} + \eta _0 \, .
\end{equation}
When the two input beams are in-phase ($\eta_0 = 0 $) and identical ($\kappa_0 = \kappa_1$), they intersect in-phase ($\Delta \varphi =0$), and so they merge into a strong central beam, see Fig.~\ref{fig:s1q2_interaction}(a). If we introduce a $1.25\%$ change in the propagation constant of the second beam ($\kappa_1 = \kappa_0 + 0.1$), then by \eqref{eq:phase_dif}, $\Delta \varphi \approx~0.1\cdot 14.7 \approx~0.48 \pi$. This phase difference is sufficient for the two beams to repel each other, see Fig.~\ref{fig:s1q2_interaction}(b). Therefore, the interaction is \textit{"chaotic"}, in the sense that a small change in the input beams leads to a large change in the interaction pattern. To further demonstrate that the change in the interaction pattern is predominately due to the phase difference, we "correct" the initial phase of the second beam by setting $\eta _0 \approx -0.48 \pi$, so that $\Delta \varphi \approx 0$ at $(z_{\rm cross}, x_{\rm cross} )$, and indeed observe that the two beams merge, see Fig.~\ref{fig:s1q2_interaction}(c)~\footnote{Unlike Fig. \ref{fig:s1q2_interaction}(a), the output beam is slightly tilted upward, since the lower input beam is more powerful, and therefore the net linear momentum points upward.}.

\begin{figure}[h!]
\centering
{\includegraphics[scale=0.53]{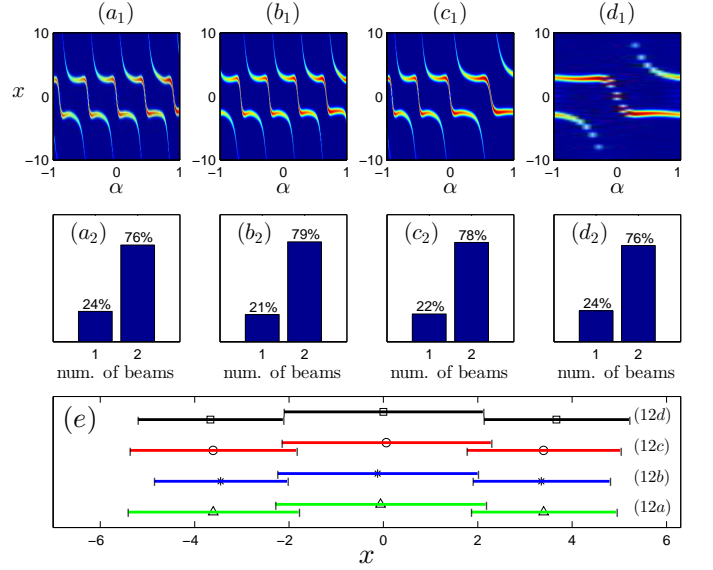}}
\caption{Solutions of the 1d cubic-quintic NLS \eqref{eq:qc_nls} with $\epsilon = 2\cdot 10^{-2}$. ($a_1$)--($d_1$): the exit intensity $|\psi(z_f=17,x;\alpha)|^2$, ($a_2$)--($d_2$): the probability of the number of output beams, and (e) the mean ($\triangle,\star, \circ,\square $) and standard deviation of the lateral location of the output beams, for the noisy initial conditions \eqref{eq:rnd_freq}--\eqref{eq:rnd_phase}, respectively. Here $a=12$, $\theta = \frac{\pi}{8}$, and $\kappa _0=8$.}
\label{fig:inter_pattern}
\end{figure}

In what follows, we consider interactions between the two crossing beams $$\psi _0 (x) = e^{i\theta x} R_{\kappa _0} (x-a) + c\cdot e^{i\eta _0 } e^{-i\theta x} R_{\kappa_1} (x+a) \, $$ with four different noise sources:
\begin{subequations}\label{eq:noise_scenarios}
\begin{align}
&\mbox{\bf random}~{\bf \kappa}: ~ \eta _0=0,~ c=1,~   \kappa_1 = \kappa_0 \left( 1+ \frac{\alpha}{8} \right) \, ,\label{eq:rnd_freq} \\
&\mbox{\bf random }\kappa\mbox{\bf , out of phase}: \, \eta _0=\pi  , \, c=1 \,,  \label{eq:off_phase_freq} \\
&\qquad \qquad \qquad \qquad \qquad \quad \quad ~\,  \kappa_1 = \kappa_0 \left( 1+ \frac{\alpha}{8} \right) \, , \nonumber \\
&\mbox{\bf random power:} ~   \eta _0= 0  ,\, c= 1+ 0.1 \alpha   , \, \kappa_1 = \kappa_0 \, , \label{eq:rnd_power}\\ 
&\mbox{\bf random phase:} ~ \eta _0= \pi \alpha   ,~  c= 1 ,~   \kappa_1 = \kappa_0  \, , \label{eq:rnd_phase}
\end{align}
\end{subequations}
where $\alpha \sim U(-1,1)$. Fig. \ref{fig:inter_pattern}(${\text a_1}$)--(${\text d_1}$) shows the "exit intensity" $|\psi(z_f, x;\alpha)|^2$ at $z_f = 17$ as a function of $x$, for $-1 \leq \alpha \leq 1$. As in Fig. \ref{fig:s1q2_interaction}, depending on $\alpha$, there are two possible outputs: Either a single beam (resulting from beam fusion), or two beams (resulting from beam repulsion). Generally speaking, there is a single output beam whenever the two input beams are "sufficiently"
in-phase at~$(z_{\rm cross},x_{\rm cross})$.%

%

In a physical setting the noise distribution is typically unknown. Nevertheless, the on-axis phase of each beam core is given by~\eqref{eq:phase_lin}, where $\kappa(\alpha)$ is a random variable. Therefore, by Lemma \ref{lem:phase_uni}, for $z_{\rm cross} \gg Z_{\rm lop}$, the phase of each beam at $(z_{\rm cross}, x_{\rm cross})$, and hence also the relative phase between them, is uniformly distributed in $[0,2\pi]$. Hence, {\em to leading order, the statistics of the interactions are universal}. Indeed, for all 4 noisy initial conditions we observe that: (i)~the probability for a single filament is $22\% \pm 2\%$ and for two filaments is $78 \% \pm 2 \%$, see Fig. \ref{fig:inter_pattern}(${\text a_2}$)--(${\text d_2}$), and (ii)~the mean and standard deviation of the lateral locations of the output beams are nearly identical, see~Fig.~\ref{fig:inter_pattern}(${\text e}$).

The above results show that the statistics of long-range interactions between laser beams are independent of the noise source and its characteristics, and can be computed using a {\em "universal model"} in which the only randomness comes from the addition of a random constant phase to one of the input beams, which is uniformly distributed in~$[0,2\pi]$, c.f. \eqref{eq:rnd_phase}. The standard approach for computing the statistics of the interactions in this "universal model" is the Monte-Carlo method. This method, however, is very inefficient due to its $O(1/\sqrt{N})$ accuracy, where $N$ is the number of NLS solutions. To efficiently use the universal model, we developed a polynomial-chaos based numerical method, which is both spectrally accurate and makes use of any deterministic NLS solver. For further details, see SM. 

In conclusion, we showed that when laser beams or pulses interact after a sufficiently long propagation distance, their relative phase at the crossing point cannot be predicted or controlled. In such cases, the notion of a "typical experiment" or a "typical solution" may be misleading, and one should adopt a stochastic approach. The loss of phase can explain some of the difficulties in phase-dependent methods in optical communications such as Quadrature Amplitude Modulation~(QAM)~\cite{skidin2016mitigation},  and in coherent combining of hundreds of laser beams in a small space for ignition of nuclear fusion~\citep{Break2001Fusion}, and for creating a more powerful laser beam \cite{mourou2013future}. In these applications, controlling the phases of the input beams or pulses might not provide a good control over their interaction or combination, due to the loss of phase. Our study suggests that controlling the relative phases at the intersection point may be achieved by either shortening the propagation distance, or by coupling the beams throughout the propagation. Loss of phase is also relevant to the loss of polarization for elliptically-polarized beams \cite{gauri2017polarization}.



\newpage

\appendix

\section{Proof of Lemma \ref{lem:phase_uni}}\label{sec:lem_proof}

For a given $z\geq 0$, denote $\varphi _z
 (\alpha) = \kappa (\alpha) +\frac{\eta _0 (\alpha)}{z} $, then $\tilde{\varphi } (\alpha ) = z\varphi_z (\alpha ) \, {\rm mod} \, (2\pi) $. We prove that $\lim\limits_{z\to \infty} \tilde{\varphi }  (\alpha ) \sim U([0,2\pi])$ by showing that for every $0\leq a <~b \leq~2\pi$, 
\begin{equation}\label{eq:lem_target_2}
\lim\limits_{z \to \infty}  \mu \left( \tilde{\varphi}  ^{-1} \left( \left[a , b \right] \right) \right) = \frac{b-a}{2\pi} \, ,
\end{equation}
where $\tilde{\varphi}  \left[a,b\right] :\,= \left\{\alpha\in [\alpha _{\min}, \alpha_{\max} ] \, | \, \tilde{u} _z (\alpha) \in [a,b] \right\}$.


We first prove the lemma for a strictly monotone function $\kappa$ on $(\alpha _{\min}, \alpha_{\max} )$. For sufficiently large $z$, $\varphi_z$ is also monotone. Let $x_k^z$ and $y_k^z$ be the solutions of
\begin{equation}\label{eq:xy_proof}
\varphi _z(x_k^z) = \frac{2k\pi +a}{z}\, ,\quad \varphi _z (y_k^z) = \frac{2k\pi +b}{z} \, ,\qquad k\in\mathbb{Z} \, .  
\end{equation}

There exists $k_{\min}  \leq k_{\max} $ such that $x^z_{k_{\min}  -1}$ and $  y^z _{k_{\max} +1 }$ do not exist, and for clarity we suppressed the dependence of $k_{\min}$ and $k_{\max}$ on $z$. By definition, $$\mu \left( \tilde{\varphi} ^{-1}  \left( \left[a , b \right] \right) \right) = \sum\limits_{k=k_{\min}}^{k_{\max}} \mu \left(x_k^z , y_k^z  \right) + E(z) =  $$
\begin{equation}\label{eq:mu_inv_exp_2}
 = \sum\limits_{k=k_{\min}}^{k_{\max}} \mu \left( \varphi_z  ^{-1}\left( \frac{2\pi k +a}{z} \right) , \varphi _z ^{-1}\left(\frac{2\pi k +b}{z}\right)  \right) +E(z) \, ,
\end{equation}
where the error term $E(z)$ exists if either $y_{k_{\min} -1} ^z  > \alpha _{\min}$ or $x_{k_{\max} +1}^z < \alpha _{\max}$ exist. In such cases, since $d\mu$ is continuous, $$E(z) = \mu\left( [\alpha _{\min},y_{k_{\min} -1}^z] \right) +\mu\left( [x_{k_{\max} +1}^z,\alpha _{\max}] \right)  \, .$$ 

We now show that if $y_{k_{\min}-1} ^z$ exists, then $\lim\limits_{z\to\infty} \mu \left( \alpha _{\min} , y_{k_{\min}-1} ^z \right) = 0$ (a similar proof holds also for $x_{k_{\max} +1} ^z$). It is enough to show that $\lim\limits_{z\to \infty} x_{k_{\min}} ^z = \alpha _{\min}$, because $y_{k_{\min -1}}^z < x_{k_{\min}}$ and $\mu$ is a continuous measure. Let $\delta >0$, then $$z\varphi _z (\alpha_{\min} +\delta) - z\varphi _z (\alpha _{\min} ) =  $$ $$ \left( \eta _0 (\alpha _{\min} +\delta) - \eta_0 (\alpha_{\min}) \right) + z\left(\kappa (\alpha _{\min}+\delta ) - \kappa _{\min} \right) $$ goes to infinity as $z\to \infty$. Therefore, for large enough $z$, $x_{k_{\min}} ^z \in (\alpha_{\min},\alpha_{\min} +~\delta)$. Thus, for all $\delta >0$, $$\alpha _{\min} \leq \lim\limits_{z\to\infty} x_{
k_{\min} }^z<\alpha _{\min} + \delta \, .$$

Since $\varphi_z$ is strictly monotone, then by the inverse function theorem $\varphi_z ^{-1}\in~C^1$, and so by substituting $\alpha = \varphi_z ^{-1}(y)  $, $$\mu\left( \varphi _z ^{-1}\left( \frac{2\pi k +a}{z}\right),\varphi _z ^{-1}\left(\frac{2\pi k +b}{z}\right) \right) =  $$ $$ \int\limits_{\varphi _z ^{-1}\left( \frac{2\pi k +a}{z}\right)} ^{\varphi _z ^{-1}\left(\frac{2\pi k +b}{z}\right)} c(\alpha) \, d\alpha \ = \int\limits_{\frac{2\pi k +a}{z}}^{\frac{2\pi k +b}{z}}g_z(y)\, dy \, \, ,$$
where $g_z(y) :\, = c(\varphi _z ^{-1} (y))(\varphi _z^{-1} )'(y)$. By Lagrange mean-value theorem, for each index $k$, there exists $\xi_k ^z \in (a,b)$ such that $$\mu\left(\varphi _z ^{-1}\left( \frac{2\pi k +a}{z}\right),\varphi _z ^{-1}\left(\frac{2\pi k +b}{z}\right) \right) =  $$ $$g_z\left(\frac{2\pi k + \xi_k ^z}{z} \right) \frac{b-a}{z} \, .$$
Substituting the above into \eqref{eq:mu_inv_exp_2} yields 
\begin{equation}\label{eq:mu_inv_exp2_2}
\mu \left( \varphi _z^{-1} \left( \left[ a , b \right] \right) \right) = \frac{b-a}{z}\sum\limits_{k=k_{\min}}^{k_{\max}} g_z\left( \frac{2\pi k + \xi_k ^z}{z} \right) + E(z) \, .
\end{equation}
Next, consider the integrals
\begin{subequations}
\begin{equation}\label{eq:inv_int_2}
I :\,= \int\limits_{\alpha _{\min }} ^{\alpha _{\max }} c\left( \alpha \right)\, dy = \mu \left( \alpha _{\min}, \alpha _{\max} \right) =1 \, , \quad 
\end{equation}
$$ I_2 :\,=\int\limits_{x_{k_{\min}} ^z} ^{y_{k_{\max}} ^z } c (\alpha) \,  d\alpha  = \int\limits^{\frac{2\pi k_{\max} +b}{z}}_{\frac{2\pi k_{\min} +a}{z}} g_z(y) \, dy  \, . $$
Using Riemann sums
\begin{equation}\label{eq:riemann_sums_2}
I_2 =  \frac{2\pi}{z} \sum\limits_{k=k_{\min}}^{k_{\max}} g_z\left(\frac{2\pi k + \xi_k ^z}{z} \right)   + O\left(z^{-1}\right) \, .
\end{equation}
\end{subequations} 
Denoting $\varphi _z(\alpha _{\min}) :\, =\varphi_{z,\min}$ and $\varphi _z (\alpha_{\max}):\,=\varphi_{z, \max}$ Since $$\int\limits_{\varphi_{z, \min}} ^{\varphi_{z, \max}} = \int\limits_{\varphi_{z, \min}} ^{\varphi _z (x_{k_{\min}}^z)} +\int\limits_{\varphi _z (x_{k_{\min}}^z)} ^{\varphi _z (y_{k_{\max}}^z)} + \int\limits_{\varphi _z (y_{k_{\max}}^z)} ^{\varphi_{z, \max}} \, , $$ then $I=I_2 + O(z^{-1})$. Equating \eqref{eq:riemann_sums_2} and  \eqref{eq:inv_int_2}, and substituting into \eqref{eq:mu_inv_exp2_2}, yields $$\mu \left( \varphi_z ^{-1}  \left( \left[ a , b \right] \right) \right) = \frac{b-a}{2\pi}  + o(1) \, ,$$ by which we prove
\eqref{eq:lem_target_2}

Finally, if $\kappa$, hence $\varphi _z$ is piece-wise monotone, we apply the above proof for each sub-interval over which $\varphi _z$ is monotone, and by additivity of measure have the result.

%
%

\section{Numerically solving the universal model}

Although in the universal model the noise is uniformly distributed, we allow for a more general noise distribution, so that we can e.g., produce results such as figure \ref{fig:lop_gpc} for non-uniform noise distributions.

Let $\psi(z,x;\alpha )$ be the solution of the NLS \eqref{eq:nls_gen} with the random initial condition~\eqref{eq:gaus_rnd_ic}. In what follows, we introduce an efficient numerical method for   computing the statistics of $g(\psi)$, e.g., the average intensity over many shots $\mathbb{E}_{\alpha}[ \left| \psi \right|^2 ]$.

The standard numerical method for this problem is Monte-Carlo, in which one draws $N$ random values of $\alpha$ and approximates $E_{\alpha} \left[ g(\alpha ) \right] \approx~\frac{1}{N} \sum\limits_{n=1}^N g(\alpha _n)$. The main drawback of this method is its slow $O(1/\sqrt{N})$ convergence rate, where $N$ is the number of NLS simulations.
If $g(\alpha): \, = g(\psi (\cdot ; \alpha) )$ is smooth in $\alpha$, however, we can use orthogonal polynomials as a \textit{spectrally accurate} basis for interpolation \cite{canuto1982approx} and numerical integration. Let $\alpha$ is distributed in $[\alpha _{\min} ,\alpha _{\max}]$ according to a PDF $c(\alpha)$, and let $\left\{ p_n (\alpha)\right\}_{n=0}^{\infty}$ be the corresponding sequence of orthogonal polynomials, in the sense that $\int\limits_{\alpha_{\min}}^{\alpha_{\max}} p_n(\alpha)p_m(\alpha) c(\alpha) \, d\alpha = \delta _{n,m}$. For example, if $\alpha$ is uniformly distributed in $[-1,1]$, then $\{p_n\}$ are the Legendre polynomials, and if $\alpha$ is normally distributed in $(-\infty, \infty)$, then $\{p_n\}$ are the Hermite polynomials. Recall that for smooth solutions one has the spectrally accurate quadrature formula $E_{\alpha} \left[ g(\alpha ) \right] \approx \sum\limits_{j=1}^{N} g(\alpha _j^N) w_j ^N\, ,$ where $\{ \alpha _j ^N \}_{j=1}^N$ and $\left\{ w_j ^N \right\}_{j=1}^{N}$ are the roots of the orthogonal polynomial $p_N(\alpha)$ and their respective weights $w_j ^N = \int\limits_{\alpha _{\min} }^{\alpha _{\max}} \prod\limits_{i=1,\, i \neq j}^N \frac{\alpha - \alpha_i ^N}{\alpha _j ^N - \alpha _i ^N} \, c(\alpha )\,d\alpha$ \footnote{See \cite{day2005roots} for a numerically efficient and stable algorithm for computing the $\left\{\alpha _j ^N, \, w_j ^N \right\}_{j=1} ^N$.}.
We apply the collocation Polynomial Chaos Expansion (PCE) method as follows~\cite{o2013polynomial,xiu2010numerical}:
\begin{framed}
\begin{enumerate}[labelindent=0pt]
\item For $j=1, \dots ,N$, solve the NLS for  $\psi \left(z, {\bf x}  ; \alpha _j ^N \right)$, and set $g(\alpha _j ^N) := g\left( \psi (z,x;\alpha_j ^N) \right)$.

\item Approximate
\begin{subequations}\label{eq:gpc_interp}
\begin{equation} 
g (\alpha ) \approx g  _N (\alpha ) :=  \sum\limits_{n=0}^{N-1} \hat{g}_N (n) p_n(\alpha ),
\end{equation}
where
\begin{equation}
\hat{g}_N (n ) = \sum\limits _{j=1} ^N p_n (\alpha _j ^N ) g \left( \alpha _j^N \right) w_j ^N \, ,\qquad n=0,\ldots , N-1 \, .
\end{equation}
\end{subequations}
\end{enumerate}
\end{framed}

This method is "non-intrusive", i.e., it does not require any changes to the deterministic NLS solver. Moreover, the orthogonality of $\{p_n \}$ leads to direct formulae for the mean and standard deviation of $g$:
$$ \mathbb{E}_\alpha \left[ g(\alpha ) \right] \approx \frac{1}{p_0}\hat{g}_N(0) \, ,$$ $$\
\sigma \left[ g( \alpha )\right]   \approx \sqrt{\sum\limits_{n=0}^{N-1} \left| \hat{g }_N(n ) \right| ^2  - \frac{\left |\hat{g }_N(0 ) \right|^2}{p_0^2}} ~ .$$

As noted, the PCE method has a spectral convergence rate for smooth functions. For example, the results in Fig. \ref{fig:lop_gpc} \ref{fig:lop2d} were computed using $N=10$ and $N=31$ NLS simulations, respectively. To reach a similar accuracy with the Monte Carlo method would require more than $1000$ NLS simulations. Some quantities of interest, however, such as the number of filaments (Fig. \ref{fig:inter_pattern}($a_2$)--\ref{fig:inter_pattern}($d_2$)), or the non-cumulative on-axis phase $\tilde{\varphi }={\rm arg} \, \left( \psi (z,{\bf x}=0;\alpha ) \right) \, {\rm mod} (2\pi)$, (Fig. \ref{fig:lop_gpc}($a_3$)--\ref{fig:lop_gpc}($c_3$) and Fig. \ref{fig:lop2d}($a_3$)--\ref{fig:lop2d}($c_3$)) are non-smooth. Therefore, a straightforward application of the PCE method for such quantities requires~$O\left( 10^3\right)$ simulations to converge. In such cases, we begin with stages (1)--(2) and calculate the PCE approximation \eqref{eq:gpc_interp} of the smooth function $\psi (z,{\bf x};\alpha)$  using~$\{\psi(z,{\bf x}; \alpha _j ^N) \}_{j=1}^N$ with a relatively small $N$. Then we proceed as follows:
\begin{framed}
\begin{enumerate}
\setcounter{enumi}{2}
\item Use the gPC interpolant \eqref{eq:gpc_interp} to obtain $\psi(\cdot, \tilde{\alpha} _m ) \approx \psi _N (\cdot, \tilde{\alpha} _m)$ on a sufficiently dense grid $\{ \tilde{\alpha} _m \} _{m=1}^M$, where $M\gg N$. 

\item Compute $g(\tilde{\alpha} _m) \approx g( \psi _N (\cdot, \tilde{\alpha } _m ) )$ for $m=1,\ldots, M$,

\item Compute the statistics of $g(\psi )$ using $\left\{ g( \psi _N (\cdot, \tilde{\alpha } _m ) ) \right\}_{m=1}^M$.
\end{enumerate}
\end{framed}
For example, when we computed the number of beams at $z=z_f$ in Fig.~\ref{fig:inter_pattern}, we first computed the PCE interpolant $\psi _N (z_f,x;\alpha)$ with $N=71$. Then we computed $\psi(z_f,x;\tilde{\alpha} _m ) \approx \psi _N (z_f,x;\tilde{\alpha }_m)$ for $m=1,\ldots,M= 801$.  For each $\tilde{\alpha} _m$, we count the number of filaments and used this to produce the histogram in figure~\ref{fig:inter_pattern}($a_2$)--\ref{fig:inter_pattern}($d_2$). The additional computational cost of sampling~$\psi _N$~\eqref{eq:gpc_interp}~at~$M\gg N$ grid points in step (3) is negligible compared to directly solving the NLS for $N$ times in stage (1).%

\bibliographystyle{apsrev}

\bibliography{LOP_bibSM}


\end{document}